\documentclass[twocolumn,english,aps]{revtex4}
\usepackage[T1]{fontenc}
\usepackage[latin9]{inputenc}
\usepackage{graphicx}

\makeatletter
\@ifundefined{textcolor}{}
{%
 \definecolor{BLACK}{gray}{0}
 \definecolor{WHITE}{gray}{1}
 \definecolor{RED}{rgb}{1,0,0}
 \definecolor{GREEN}{rgb}{0,1,0}
 \definecolor{BLUE}{rgb}{0,0,1}
 \definecolor{CYAN}{cmyk}{1,0,0,0}
 \definecolor{MAGENTA}{cmyk}{0,1,0,0}
 \definecolor{YELLOW}{cmyk}{0,0,1,0}
 }

\makeatother

\makeatother

\usepackage{babel}

\makeatother

\usepackage{babel}

\begin{document}

\title{Entanglement cost of two-qubit orthogonal measurements }

\author{Somshubhro Bandyopadhyay}

\affiliation{Department of Physics and Center for Astroparticle Physics and Space
Science, Bose Institute, Block EN, Sector V, Kolkata 700091, India}

\email{som@bosemain.boseinst.ac.in}

\author{Ramij Rahaman}

\email{Ramij.Rahaman@ii.uib.no}

\affiliation{Department of Informatics, University of Bergen, PB-7803, Bergen-5020,
Norway}

\author{William K. Wootters}

\affiliation{Department of Physics, Williams College, Williamstown, MA 01267,
USA \\
 Department of Applied Physics, Kigali Institute of Science and
Technology, B.P. 3900, Kigali, Rwanda}

\email{William.K.Wootters@williams.edu}

\maketitle
\noindent {\em Abstract}. The {}``entanglement cost'' of a bipartite
measurement is the amount of shared entanglement two participants
need to use up in order to carry out the given measurement by means
of local operations and classical communication. Here we numerically
investigate the entanglement cost of generic orthogonal measurements
on two qubits. Our results strongly suggest that for almost all measurements
of this kind, the entanglement cost is strictly greater than the average
entanglement of the eigenstates associated with the measurements,
implying that the nonseparability of a two-qubit orthogonal measurement
is generically distinct from the nonseparability of its eigenstates.%\textquotedbl{}

\bigskip{}

Certain measurements on composite systems, whose parts are spatially
separated, cannot be implemented by local operations and classical
communication (LOCC). Entangling measurements certainly belong to
this category, and the so-called {}``non-locality without entanglement''
shows that even some unentangled measurements cannot be performed
by LOCC alone \cite{BDFMRSSW99}. The latter example clearly exhibits
a kind of nonseparability associated with joint quantum measurements
that is not entirely captured by the entanglement of their eigenstates.

It is, however, possible to quantify this nonseparability by computing
the amount of entanglement that is required to perform a measurement
on spatially separated subsystems by LOCC. This amount is referred
to as the {}``entanglement cost'' of the given measurement, the
precise definition of which is given below.

We imagine two parties, Alice and Bob, each holding one of the two
particles to be measured. They are allowed to do any sequence of LOCC
but are not allowed to transmit quantum information. Rather we give
them, as a resource, shared entangled pure states (whose form Alice
and Bob are allowed to choose), and we keep track of the amount of
entanglement they spend in performing the measurement. For a pure
state $|\psi\rangle_{AB}$ of a bipartite system $AB$, the entanglement
is \begin{equation}
\mathcal{E}\left(|\psi\rangle_{AB}\right)=-\mbox{Tr}\left(\rho_{A}\log\rho_{A}\right),\end{equation}
 where $\rho_{A}=\mbox{Tr}_{B}\left(|\psi\rangle_{AB}\langle\psi|\right)$
is the reduced density matrix of the subsystem $A$. The logarithm
is always taken with respect to base two, so the entanglement is measured
in ebits. If Alice and Bob completely use up a copy of a state $|\eta\rangle$,
then the entanglement cost of that operation is simply $\mathcal{E}\left(|\eta\rangle\right)$.
On the other hand, if they convert an entangled state into a less
entangled state, then the cost is the difference, that is, the amount
of entanglement lost. It should be noted that a different notion of
the entanglement cost is considered in Ref. \cite{JKLPPSW03}, namely
the amount of entanglement needed to effect a Naimark extension of
a given POVM.

Computing the exact entanglement cost of a generic bipartite measurement
seems to be a hard problem except in special cases. These cases include
any maximally entangled measurement in $d\otimes d$, which costs
exactly $\log_{2}d$ ebits (a particular implementation achieving
this cost uses a maximally entangled state to teleport the information
in one of the two parts to the location of the other), and any complete
measurement in $d\otimes d$ which is invariant under all local Pauli
operations, for which the cost is equal to the average entanglement
of the states associated with the outcomes.

Notwithstanding the difficulty of computing the exact entanglement
cost of measurements, some progress has been made in obtaining lower
and upper bounds \cite{BBKW09,BHLS03,CGB01,CDKL01,Ber07,JP99,Vid99}.
A general lower bound can be computed by considering the entanglement
production capacity of measurements \cite{GKRSS01,HSSH03,Smo01}.
In this way one can show that for any complete measurement on a bipartite
system the entanglement cost is at least as great as the average entanglement
of the pure states associated with the outcomes. We call this lower
bound the {}``entropy bound,'' since the entanglement of a pure
bipartite state is quantified by the entropy of either of the two
parts. This bound is achieved for the special classes of measurements
mentioned in the preceding paragraph.

Interestingly, for a certain class of orthogonal partially entangling
measurements on two qubits, the entanglement cost has been shown to
be strictly greater than the average entanglement of the eigenstates
\cite{BBKW09}, suggesting that this feature is perhaps a generic
property of measurements. If it is, then one can say that the nonseparability
of a measurement is generically a distinct property from the nonseparability
of its eigenstates. In this paper we investigate, by means of numerical
calculations, whether this is the case for complete orthogonal measurements
on two qubits.

Our numerical results strongly indicate that for generic orthogonal
two-qubit measurements, entanglement cost is indeed strictly greater
than the average entanglement of the measurement eigenstates; that
is, it is strictly greater than the entropy bound. We reach this conclusion
by computing another lower bound, again based on entanglement production
but with a more refined analysis than the one leading to the entropy
bound. We evaluate this bound for a broad sample of orthogonal measurements,
covering fairly densely the range of possible values of the average
entanglement of the eigenstates. Though there are exceptional cases
for which the more refined bound is equal to the entropy bound, it
appears that the set of such cases is very small, probably constituting
a manifold of lower dimension than the manifold of all orthogonal
measurements. The exceptional cases that we can explicitly identify
are the following: (i) any measurement in a product basis (for which
the entanglement cost is zero), (ii) any measurement whose eigenstates
are all maximally entangled (for which the entanglement cost is $1$
ebit), and (iii) any measurement that is equivalent, under local unitaries,
to the measurement with eigenstates $\left\{ \frac{1}{\sqrt{2}}\left(|00\rangle+|11\rangle\right),\frac{1}{\sqrt{2}}\left(|00\rangle-|11\rangle\right),|01\rangle,|10\rangle\right\} $.

A complete orthogonal projective measurement $M$ is specified by
a collection of rank-one projection operators $\Pi_{i}=|\psi_{i}\rangle\langle\psi_{i}|$
that sum to the identity. Such operators are necessarily orthogonal;
that is, $\mbox{Tr}(\Pi_{i}\Pi_{j})=\delta_{ij}$. In this paper we
understand $M$ to provide only a rule for computing probabilities,
not for computing the final state of the system after measurement.
So a realization of $M$ can be any procedure that yields the correct
probabilities, even if, for example, it destroys the measured system.
We allow the possibility of probabilistic measurement procedures,
in which the probabilities might depend on the initial state of the
system being measured. As we will be computing an average cost, we
need to specify an initial state in order for the average to be well
defined. We assume that this initial state is the completely mixed
state, which we regard as the least biased choice. We now give the
formal definition of entanglement cost associated with a quantum measurement.

Given a measurement $M$, let $\mathcal{P}(M)$ be the set of all
LOCC procedures $P$ such that (a) $P$ uses pure entangled states
and LOCC and (b) $P$ realizes $M$ exactly. Then $C(M)$, the entanglement
cost of measurement $M$, is defined to be \begin{equation}
C(M)=\inf_{P\in\mathcal{P}(M)}\langle\mathcal{E}_{\mbox{\scriptsize initial}}-\mathcal{E}_{\mbox{\scriptsize final}}\rangle\end{equation}
 where $\mathcal{E}_{\mbox{\scriptsize initial}}$ is the total entanglement
of all the resource states used in the procedure, $\mathcal{E}_{\mbox{\scriptsize final}}$
is the distillable entanglement of the state remaining at the end
of the procedure, and $\langle...\rangle$ indicates an average over
all possible results of $P$, when the system on which the measurement
is being performed is initially in the completely mixed state.

As was mentioned above, our lower bound on the entanglement cost is
obtained by considering the entanglement production capacity of $M$.
That this capacity is a lower bound on the entanglement cost of $M$
follows from the fact that in performing the measurement, the participants
must consume at least as much entanglement as the measurement can
produce, since entanglement cannot increase, on an average, by LOCC.
Specifically we imagine that in addition to qubits $A$ and $B$ on
which Alice and Bob want to perform their measurement, they also have
in their possession auxiliary qubits $C$ and $D$. (Here $A$ and
$C$ are held by Alice, and $B$ and $D$ are held by Bob.) Now consider
an initial state of four qubits such that the measurement $M$ on
qubits $A$ and $B$ collapses qubits $C$ and $D$ into a possibly
entangled state. Then, the average amount by which the measurement
increases the entanglement between Alice and Bob is a lower bound
on $C(M)$. That is, \begin{equation}
C\left(M\right)\geq C_{L}\left(M\right)=\overline{\mathcal{E}}_{CD}-\mathcal{E}_{AC:BD},\end{equation}
 where, $\overline{\mathcal{E}}_{CD}$ is the average final entanglement
between the qubits $C$ and $D$ and $\mathcal{E}_{AC:BD}$ is the
initial entanglement across the bipartition $AC:BD$. We do not consider
any final entanglement between $A$ and $B$, because we want our
lower bound to apply to any procedure that implements $M$, even if
it destroys $A$ and $B$.

The initial state is chosen to be \begin{equation}
|\chi\rangle_{ABCD}=\frac{1}{2}\sum_{i=1}^{4}|\psi_{i}\rangle_{AB}\otimes|\phi_{i}\rangle_{CD}\end{equation}
 where the states $\left\{ |\psi_{i}\rangle,i=1,...,4\right\} $are
the orthogonal eigenstates of the measurement $M$ that Alice and
Bob want to perform. The states $\left\{ |\phi_{i}\rangle,i=1,...,4\right\} $
are what we will call the {}``detector'' states, so named because
they correspond to the measurement outcomes on the system $AB$. While,
for a given measurement $M$, its eigenstates are fixed, the detector
states can be completely arbitrary, except, we insist that they be
mutually orthogonal. This restriction guarantees that the initial
state of $AB$ is the completely mixed state, in accordance with our
definition of entanglement cost. For any particular choice of the
detector states, a lower bound on $C(M)$ is given by the formula
\begin{equation}
C(M)\geq C_{L}\left(M\right)=\frac{1}{4}\sum_{i=1}^{4}E(|\phi_{i}\rangle)-\mathcal{E}_{AC:BD}.\end{equation}
 The above quantity can be maximized numerically over the detector
states to obtain the best absolute lower bound as has been done in
Ref. \cite{BBKW09} for a class of two-qubit orthogonal measurements.
However, here our objective is not to get the best possible lower
bound for $M$. We only want to find out whether the lower bound is
typically greater than the entropy bound. We therefore consider the
quantity \begin{equation}
\delta=C_{L}(M)-\frac{1}{4}\sum_{i=1}^{4}E(|\psi_{i}\rangle)\end{equation}
 and maximize it only over a discrete grid of detector states. If
the resulting maximum is unambiguously positive, that is, greater
than the numerical error, we can conclude that for the given measurement
$M$, the entanglement cost is strictly greater than the entropy bound.

For a generic orthogonal two-qubit measurement $M$ its eigenstates
are simply a set of orthogonal vectors $\left\{ |\psi_{i}\rangle,i=1,2,3,4\right\} $
in $2\otimes2$. The most general canonical form of four orthogonal
states in $2\otimes2$, up to local unitaries, can be obtained in
the following fashion. Walgate \emph{et al.} \cite{WSHV00} have shown
that any pair of orthogonal states of two qubits can, by local rotations,
be brought to the form $\{\alpha|00\rangle+\beta|11'\rangle,\gamma|01\rangle+\delta|10'\rangle\}$,
where $|0'\rangle$ and $|1'\rangle$ constitute an orthogonal basis
for the second qubit. Thus we can write the first two eigenstates
of our measurement as \begin{equation}
|\psi_{1}\rangle=\cos(a)|00\rangle+e^{ib}\sin(a)|1\rangle(\cos(u)|0\rangle+e^{iv}\sin(u)|1\rangle);\end{equation}
 \begin{equation}
|\psi_{2}\rangle=\cos(c)|01\rangle+e^{id}\sin(c)|1\rangle(e^{-iv}\sin(u)|0\rangle-\cos(u)|1\rangle).\end{equation}
 (Since the quantities of interest are independent of the overall
phases of the state vectors, we are free to take the coefficient of
the first term in each of these expressions to be real and non-negative.)
We now define two states that are orthogonal to both $|\psi_{1}\rangle$
and $|\psi_{2}\rangle$: \begin{equation}
|\psi_{1}^{\perp}\rangle=e^{-ib}\sin(a)|00\rangle-\cos(a)|1\rangle(\cos(u)|0\rangle+e^{iv}\sin(u)|1\rangle);\end{equation}
 \begin{equation}
|\psi_{2}^{\perp}\rangle=e^{-id}\sin(c)|01\rangle-\cos(c)|1\rangle(e^{-iv}\sin(u)|0\rangle-\cos(u)|1\rangle).\end{equation}
 The remaining two eigenstates of the measurement will be linear combinations
of $|\psi_{1}^{\perp}\rangle$ and $|\psi_{2}^{\perp}\rangle$: \begin{equation}
|\psi_{3}\rangle=\cos(x)|\psi_{1}^{\perp}\rangle+e^{iy}\sin(x)|\psi_{2}^{\perp}\rangle;\end{equation}
 \begin{equation}
|\psi_{4}\rangle=e^{-iy}\sin(x)|\psi_{1}^{\perp}\rangle-\cos(x)|\psi_{2}^{\perp}\rangle.\end{equation}
 We cover the full set of orthogonal measurements by allowing the
following ranges for the parameters: $0\le a,c,u,x\le\pi/2$, $0\le b,d,v,y\le2\pi$.

To generate the specific subset of orthogonal measurements to be considered
in our numerical calculation, we step through the ranges of the parameters,
using the following step sizes: for $a,c$, and $u$, step size $\pi/24$;
for $b,d,v$, and $y$, step size $\pi/12$; for $x$, step size $\pi/16$.
For any given measurement $M$, we parameterize the detector states
$\{|\phi_{i}\rangle,i=1,...,4\}$, in the same way using a different
set of parameters, and use step sizes half as large as those we use
for the measurement states. In each case, that is, for each measurement,
we record the largest value of $\delta$ obtained by stepping through
the discrete grid of detector states. %
\begin{figure*}
\centerline{ \mbox{\includegraphics[width=5in]{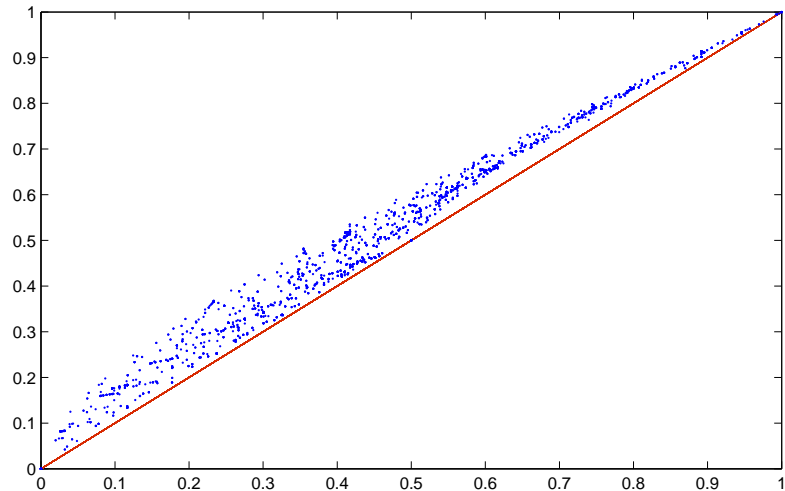}} % \mbox{\includegraphics[width=3.00in]{finalcost.eps}}
 } 

\caption{Blue dots: Plot of the lower bound of the entanglement cost $C_{L}(M)$
with respect to the entropy bound for general two-qubit orthogonal
measurements. Red line: Plot of the entropy bound.}

\label{fig:finalcost} 
\end{figure*}

Our results are plotted in Fig.~\ref{fig:finalcost}. (The plot includes
some additional points not covered by the grid described above. These
points were chosen more or less arbitrarily.) For almost every measurement
we considered, the refined lower bound is strictly larger than the
entropy bound. The only exceptions we have found are those mentioned
earlier: the cases of product-state or maximally entangled measurements,
and the one intermediate case, with eigenstates $\left\{ \frac{1}{\sqrt{2}}\left(|00\rangle+|11\rangle\right),\frac{1}{\sqrt{2}}\left(|00\rangle-|11\rangle\right)|01\rangle,|10\rangle\right\} $.

Our results strongly suggest that for almost all orthogonal two-qubit
measurements, the lower bound on the entanglement cost is strictly
greater than the average entanglement of the states themselves. Thus
the nonseparability associated with a measurement appears to generically
exceed the nonseparability of its eigenstates. We note that many of
the points plotted in the lower half of Fig.~\ref{fig:finalcost}
lie very close to the entropy bound. This fact suggests that there
may be a class of special measurements, which we have not yet identified
explicitly, for which our lower bound, even when maximized over all
detector states, is exactly equal to the entropy bound. It would be
interesting to identify such measurements if they do indeed exist.

Note that the lower bounds are also valid asymptotically. Suppose
$N$ pairs of qubits are given to Alice and Bob and they are to perform
the same measurement on each pair. It is conceivable that by performing
a measurement involving all $N$ pairs they can achieve better efficiency.
Even in this setting the lower bounds obtained are still applicable.
To see this, imagine that each pair is initially entangled with a
pair of auxiliary qubits. Since both the initial and final entanglements
of the whole system across the bipartition $AC:BD$ are proportional
to $N$, the original argument still holds.

Several open questions remain. The method presented here and in Ref.\cite{BBKW09}
works reasonably well for obtaining lower bounds, although in higher
dimensions numerical calculations could become much more complex.
Obtaining {\em upper bounds}, even for two-qubit measurements,
is still very much an open question. Only for special classes of orthogonal
measurements have non-trivial upper bounds been obtained, and even
then only for a particular range of entanglement of the eigenstates.
Pursuing these questions should shed more light on the nature of nonseparability
in quantum measurements.
\begin{acknowledgments}
R.R. would like to acknowledge thankfully the financial support of
the Norwegian Research Council. 

%ramij_0

\end{acknowledgments}

\end{document}